\documentclass{cflowproc}

\usepackage{natbib}

\begin{document}


\shortauthors{Gitti, Myriam et al.}     
\shorttitle{Cooling Flows and Radio Mini--Halos} 

\title{Cooling Flows and Radio Mini--Halos in Clusters of Galaxies}   

\author{Myriam Gitti,\affilmark{1,2,3}   
Gianfranco Brunetti,\affilmark{3}                
Giancarlo Setti \affilmark{2,3} and Luigina Feretti\affilmark{3}}

\affil{1}{Institute of Astrophysics, University of Innsbruck,
Technikerstra{\ss}e 25, A-6020 Innsbruck, Austria}   
\and                                
\affil{2}{Dipartimento di Astronomia, Universit\`a di Bologna,
via Ranzani 1, I--40127 Bologna, Italy }
\and
\affil{3}{Istituto di Radioastronomia del CNR,
via Gobetti 101, I--40129 Bologna, Italy}


\begin{abstract}
We apply to radio mini--halo candidates a model for electron 
re--acceleration in cooling flows (\citeauthor*{gbs} 2002).
In agreement with the basic idea of the model, we find that the 
power required for the re--acceleration of the electron population is only
a small fraction of the maximum power that can be extracted from the cooling 
flow (as estimated on the basis of the standard model). 
Observationally, we notice that the strongest radio mini--halos are found
in association with the most powerful cooling flows, and that cooling flow 
powers are orders of magnitude larger than the integrated radio power.
Possible effects of new {\it Chandra} and {\it XMM--Newton} estimates of 
$\dot{M}$ on this trend are considered: we conclude that even if
earlier derived cooling rates were overestimated, cooling flow powers  
are still well above the radio powers emitted by mini--halos.
\end{abstract}



\section{Introduction}
\label{Gitti:intro.sec}

It is well known that the radio emission from clusters of galaxies 
generally originates from individual radio emitting galaxies.
In addition, some clusters of galaxies show synchrotron emission not 
directly associated with a particular galaxy but rather diffused into the 
intra--cluster medium (ICM). 
These diffuse radio sources, which are the most spectacular probe of the 
existence on non--thermal emission from the ICM,
have been classified in three classes: 
cluster--wide halos, relics and mini--halos \citep{feretti96}.
Both cluster--wide halos and relics have low surface brightness, large
size ($\sim$ 1 Mpc) and steep spectrum, but the former are located at the 
cluster centers and show low or negligible polarized emission, while
the latter are located at the cluster peripheries and are highly
polarized. 
The diffuse radio emission is found in clusters which show significant 
evidence for an ongoing merger (e.g., \citeauthor{edge92} 1992; 
\citeauthor{giovannini02} 2002). 
It was proposed that recent cluster mergers may play an 
important role in the re--acceleration of the radio--emitting relativistic 
particles, thus providing the energy to these extended sources
(e.g., \citeauthor{tribble93} 1993; \citeauthor{brunetti01} 2001).
The merger picture is consistent with the occurrence of large--scale 
radio halos in clusters without a cooling flow, since the major merger 
event is expected to disrupt a cooling flow (e.g., \citeauthor{sarazin02}
2002 and references therein).

Mini--halos are diffuse radio sources, extended on a smaller scale
(up to $\sim 500$ kpc), surrounding a dominant radio galaxy at the cluster 
center.  
They are only observed in clusters with a cooling flow and it has been found
that these sources are not connected to ongoing cluster merger activity.
Their radio emission is indicative of the presence of diffuse relativistic
particles and magnetic fields in the ICM, 
since these sources do not  appear as extended
lobes maintained by an Active Galactic Nucleus (AGN), as in classical radio 
galaxies \citep*{giovannini02}.

More specifically, \citet*[hereafter GBS]{gbs} suggested 
that the diffuse synchrotron emission from radio mini--halos 
is due to a relic 
population of relativistic electrons re--accelerated by MHD turbulence 
\textit{via} Fermi--like processes, with the necessary energetics 
supplied by the cooling flow. 
Very recently, the alternative possibility of hadronic origin for 
radio mini--halos has also been discussed \citep{pfrommer03}.
 
Here, we present the application of GBS's model to radio mini--halo candidates
and discuss the observational properties of the population of radio 
mini--halos.

In particular, in Sect. 2 we briefly review GBS's model,
in Sect. 3 we present the application of the model to the Perseus cluster
and to the new mini--halo candidate A2626, and discuss the results.
In Sect. 4 we present the observational properties of radio mini--halo 
candidates in relation to those of host clusters.

A Hubble constant
$\mbox{H}_0 = 50 \mbox{ km s}^{-1} \mbox{ Mpc}^{-1}$ 
is assumed.
The radio spectral index $\alpha$ is defined such as 
$S_{\nu} \propto \nu^{-\alpha}$ and,
where not specified, all the formulae are in cgs system.


\section{Model for the Origin of Radio Mini--Halos}
\label{Gitti:mymodel.sec}

The spatial extension of the diffuse radio emission from mini--halos
represents a serious problem for understanding the origin of these sources.
The radiative lifetime of an ensemble of relativistic electrons 
losing energy by synchrotron emission and inverse Compton (IC) scattering off 
the CMB photons is given by:
\begin{equation}
\label{Gitti:tau_losses.eq}
\tau_{\rm sin+IC} \simeq \frac{24.5}{[(B^2 + B_{\rm \tiny CMB}^2) \, \gamma]}
 \; \; \; \mbox{yr}
\end{equation}
where $B$ is the magnetic field intensity (in G), $\gamma$ is the Lorentz 
factor and $B_{\rm \tiny CMB} = 3.18 \times 10^{-6} (1+z)^2$ G is the 
magnetic field equivalent to the CMB. 
In a cooling flow region (i.e. for distances $r < r_{\rm c}$, the cooling 
radius) the compression of the thermal ICM is expected to produce a 
significant 
increase of the strength of the frozen--in cluster magnetic field
and consequently of the electron radiative losses.
Therefore, in the absence of a re--acceleration or continuous injection 
mechanisms, relativistic electrons injected at a given time in these strong 
fields (of order of 10 - 20 $\mu$G, \citeauthor{ge93} 1993; 
\citeauthor{carilli02} 2002) 
should already have lost most of their energy and the radio emission would 
not be observable for more than $\sim 10^7$ - $10^8$ yr.
Such a radiative lifetime is orders of magnitude shorter than the 
characteristic diffusion time necessary for the radio--emitting electrons
to cover the distance on which the diffuse radio emission from mini--halos
is observed, therefore it is plausible that some re--acceleration mechanisms 
able to balance the radiative losses and modify the electron energy spectrum 
is acting.

In order to evaluate the radiative losses in the cooling
flow region at any distance from the cluster center it is necessary 
to parameterize the radial dependence of the field strength, 
which depends on the compression of the thermal gas in the cooling
region (i.e., on $n(r/r_{\rm c})$, $r_{\rm c}$ being the
cooling radius).
On the one hand, the X--ray emission and the spectra of cooling flow clusters 
unavoidably show the increase of the gas density and the decrease
of the temperature towards the center of these clusters.
On the other hand, {\it Chandra} and {\it XMM--Newton}
observations have pointed out that the mass deposition rates
in cooling flows have been overestimated in the
past by a factor 5-10 \citep[e.g.,][]{fabian03};
the new mass deposition rates are not too
different from the masses of cold gas estimated by 
the recent IRAM observations (\citeauthor{edge01} 2001; \citeauthor{salome03}
2003).
In addition, high spectral resolution
observations with the Reflection Grating Spectrometer (RGS)
on {\it XMM--Newton} do not show evidence for gas cooling
at temperature lower than 1-2 keV \citep[e.g.,][]{peterson03}.

All these studies basically confirm the idea that the
gas cools and thus it results compressed in the central region
of the cooling cores, but on the other hand they also
point out that the physics of the ICM in these regions
is more complex than that described by the standard cooling flow 
models.
Unfortunately, no successful model 
in alternative to the standard model
has been proposed yet, therefore, 
as a first approximation in our model calculations, 
we will evaluate the
radial behaviour of the physical quantities in the ICM
making use of the standard - single phase - cooling flow model.

In the framework of this model, the 
increase of the strength of the frozen--in cluster magnetic field
is 
$B \propto r^{-2}$ for radial compression \citep{soker90}
or $B \propto r^{-0.8}$ for isotropic compression \citep{tribble93}.

The time evolution of the energy of a relativistic electron in these regions
is determined by the competing processes of losses and re--accelerations 
(both related to the magnetic field) :
\begin{equation}
\dot{\gamma}(x) = - \beta(x) \gamma^2(x) + \alpha_+(x) \gamma(x) - \chi(x) 
\label{Gitti:gammapunto.eq}
\end{equation}
where $x=r/r_{\rm c}$,
$\beta$ is the coefficient of synchrotron and IC losses, 
$\alpha_+$ the re--acceleration coefficient and $\chi$ the Coulomb loss term.

The main assumption of GBS's model is that a relic population of 
relativistic electrons (originated for example in a central active 
galaxy and then spread throughout the whole cooling flow region
by diffusion in few Gyr)
as well as a seed large--scale turbulence (naturally originated in the ICM 
for example as a result of past merger events 
or motion of the galaxies in the cluster, or due to the interaction of the 
radio lobes of the central galaxy with the cooling flow itself),
are present in the cooling flow region, and that the relic electrons
can be efficiently re--accelerated via Fermi--like processes 
by magnetohydrodynamic (MHD) turbulence amplified by the compression of 
the magnetic field in the cooling flow region. 
However, the turbulence must not be too high in order to avoid the disruption 
of the cooling flow: this fine--tuning of the turbulent energy could explain, 
at least qualitatively, the rarity of these radio sources.

In the present paper, we consider a coefficient for systematic 
re--acceleration given by \citep{melrose80}:
\begin{equation} 
\label{Gitti:coeffermi.eq}
\alpha_+(x) \approx (\pi/c) \, v_{\rm A}^2(x) \, l^{-1}(x) \,
[\delta B(x) / B(x)]^2 
\end{equation}
where $v_{\rm A} = \sqrt{B^2/4 \pi \rho}$ is the Alfv\'en velocity, and $l$ 
is the characteristic MHD turbulence scale.
For simplicity we assume $\delta B(x) / B(x) = {\rm const}$.
The energy at which the losses are balanced by the re--acceleration, 
 $\gamma_{\rm b}$,  
is obtained by Eq. \ref{Gitti:gammapunto.eq} and, 
since Coulomb losses are nearly negligible at such electron energies, it is 
$\gamma_b \approx \alpha_{+}/\beta$. 

Under these assumptions, the stationary spectrum of the relativistic 
electrons is given by \citep{gitti03}: 
\begin{equation}
\label{Gitti:ennegamma.eq}
N(\gamma, x) \approx N(\gamma_{\rm b})_{\rm c} 
\left({{\gamma_{{\rm b,c}} }\over
{\gamma_{\rm b}(x)}}\right)^2  {\rm e}^2 x^{-s} \cdot 
\left({{\gamma}\over{\gamma_{\rm b}(x)}}\right)^2 
{\rm e}^{(-2 \gamma/\gamma_{\rm b}(x))} 
\end{equation}
which is essentially peaked at $\gamma_{\rm b}$ and where 
the electron energy density has been parameterized as 
$\propto x^{-s}$, $s$ being a free parameter.
The other free parameters are 
$B_{\rm c}=B(r_{\rm c})$ and $l_{\rm c}=l(r_{\rm c})$.


\section{Model application to observed mini--halos}
\label{Gitti:application.sec}

\begin{figure*}
\plotone{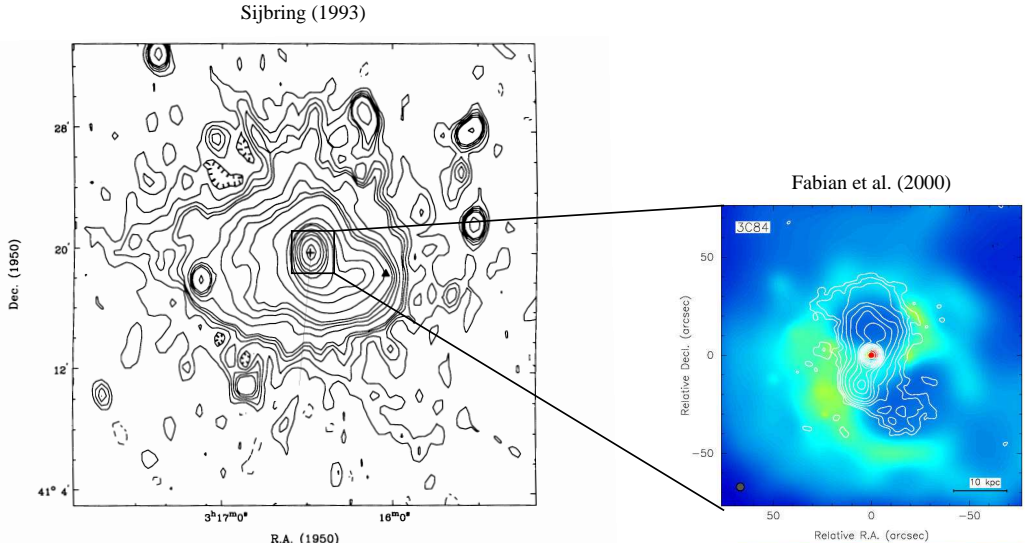}
\figcaption{\textbf{~Left panel}: 327 MHz map of the mini-halo in the Perseus 
cluster at a resolution of $51'' \times 77''$ (Sijbring 1993). 
The cross indicates the position of the cD galaxy NGC 1275. 
\textbf{~Right panel}: X--ray (grey scale)/radio (contours) overlay for the 
central part of the Perseus cluster around NCG 1275 (Fabian et al. 2000).
X--ray data are obtained with {\it Chandra}.
From \citet{gitti03}.
\label{Gitti:perseus.fig}}
\end{figure*}

In this Section we will present the application of GBS's model to 
the prototype of the mini--halo source class, the Perseus cluster,
and a new mini--halo candidate, Abell 2626, and 
discuss the physical implications of the results.


\subsection{Observed mini--halos}

\subsubsection*{The Perseus Cluster}

\begin{figure}
\plotone{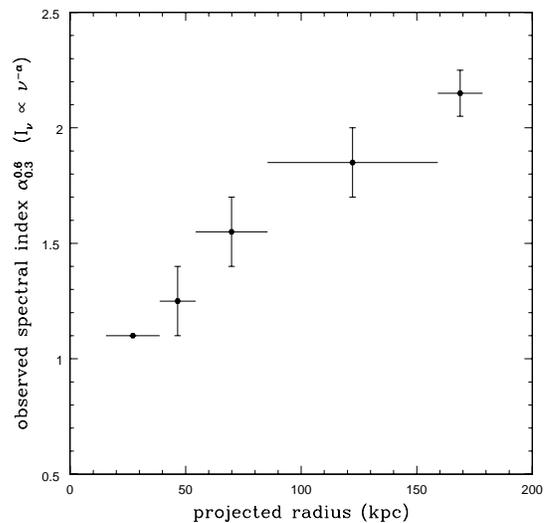}
\figcaption{
Radial spectral steepening observed in Perseus 
along the direction indicated in Fig. \ref{Gitti:perseus.fig}.
Radio data are taken from \citet{sijbring93}.
\label{Gitti:perseus_irr.fig}}
\end{figure}

We believe that
the most striking evidence for electron re--acceleration in cooling flows is
the case of the Perseus cluster (A426, z=0.0183).
The diffuse radio emission from the mini--halo 
(see left panel in Fig. \ref{Gitti:perseus.fig})
has a total extension of $\sim 15'$ (at the redshift of the cluster 
$1'$ corresponds to $\sim 30$ kpc) 
and its morphology is correlated with that of the cooling flow X-ray map
(\citeauthor{sijbring93} 1993; \citeauthor{ettori98} 1998).
Moreover, the mini--halo show a very steep spectrum 
(the integrated spectral index between 
$\nu=327$ MHz and $\nu = 609$ MHz is $\alpha \sim 1.4$, 
\citeauthor{sijbring93} 1993)
which steepens with distance from the center
(see Fig. \ref{Gitti:perseus_irr.fig}).

On smaller scales ($\sim 1'$), there is evidence of interaction between 
the radio lobes of the central radiogalaxy 3C84 and the 
X--ray emitting intra--cluster gas (e.g., \citeauthor{bohringer93} 1993; 
\citeauthor{fabian00} 2000; see Fig. \ref{Gitti:perseus.fig}, right panel).
A recent interpretation is that the holes in the 
X--ray emission are due to buoyant radio lobes 
which are currently expanding subsonically 
\citep{fabian02}.  
It is important to notice that 
the spectral index in these lobes ranges from $\sim 0.7$
in the center to $\sim 1.5$ in the outer regions \citep{pedlar90}, 
which is a value similar to the spectrum of the mini--halo
extended over a scale $\sim 10$ times larger.  
Thus it is difficult to find a direct connection between the radio lobes and 
the mini--halo in terms of simple particle diffusion or buoyancy as,
in this case, the diffusion time is considerably greater than the
radiative lifetime of the radio electrons, while adiabatic
losses during the expansion
would cause a very rapid steepening of the spectrum preventing
the detection of large--scale radio emission.

Thus, by assuming a primary origin for relativistic electrons in 
radio mini--halos,
efficient re--acceleration mechanisms in the cooling flow 
region are necessary to explain the presence of the
large--scale radio emission in Fig. \ref{Gitti:perseus.fig}.

\subsubsection*{Abell 2626}

\begin{figure}
\plotone{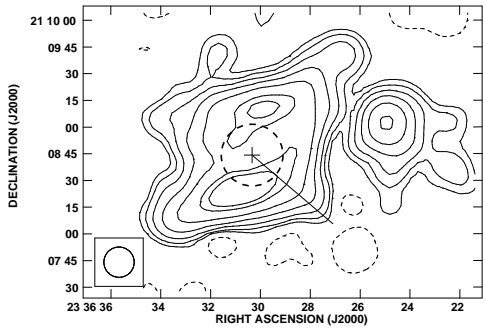}
\figcaption{
1.5 GHz VLA map of A2626 at a resolution of $17'' \times 17''$
after the subtraction of the central discrete source indicated by the cross. 
The contour levels are $-0.06$ (dashed), 0.06, 0.12, 0.24, 0.48, 0.96, 1.92,
2.5, 4, 8, 10, 12 mJy/beam. The r.m.s. noise is 0.02 mJy/beam.
The dashed line defines the region in which GBS's model is not applicable,
while the solid line represents the direction we have considered for the
surface brightness profile (see Fig. \ref{Gitti:brillanza.fig}).
\label{Gitti:a2626halo.fig}}
\end{figure}

The cluster A2626 hosts a relatively strong cooling flow 
\citep{white97} and contains an amorphous radio 
source near to the center (\citeauthor{roland85} 1985; \citeauthor{burns90}
1990) which is extended on a scale comparable to that of the cooling flow 
region with an elongation coincident with the X--ray distribution 
\citep{rizza00}.

In order to extend the application of GBS's model to this cluster, 
we have requested and analyzed some of the VLA archive data 
of A2626 with the aim to derive the surface 
brightness map, the total spectral index and the spectral index distribution 
of the diffuse radio emission. 
Standard data reduction was done using the National Radio Astronomy 
Observatory (NRAO) AIPS package.
In particular, we produced a 1.5 GHz map of the diffuse radio emission 
after subtraction of the central discrete source 
(Fig. \ref{Gitti:a2626halo.fig}).
To be conservative, the region in which the surface brightness
is affected by the subtraction procedure (within the dashed circle 
in Fig. \ref{Gitti:a2626halo.fig}) has not been considered in the application 
of GBS's model. 

\begin{figure}
\plotone{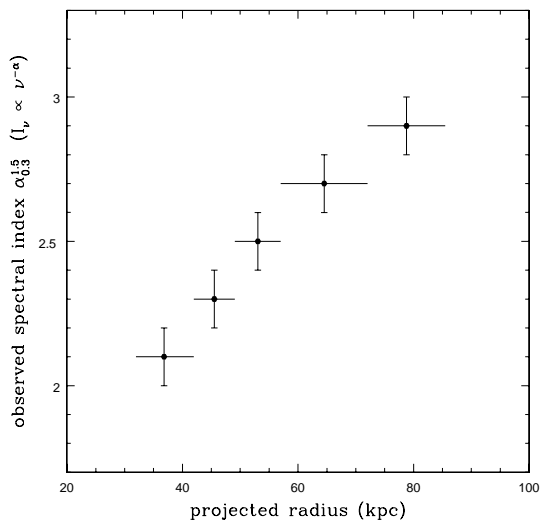}
\figcaption{
Radial spectral steepening observed in A2626
along the direction indicated in Fig. \ref{Gitti:a2626halo.fig}.
\label{Gitti:a2626_irr.fig}}
\end{figure}

The integrated spectral index of the diffuse emission between 
$\nu=330$ MHz and $\nu = 1.5$ GHz is $\alpha \sim 2.4$. 
By studying the spectral index distribution we found that,
as in the case of Perseus, the diffuse radio emission shows a steep spectral
index which steepens with the distance from the center 
(see Fig. \ref{Gitti:a2626_irr.fig}).

The radio results are summarized in Table \ref{Gitti:risradio.tab}
(see \citeauthor{gitti03} 2003 for a more detailed discussion).

\begin{table}
\begin{center}
\caption{Radio results for A2626}
\begin{tabular}{ccccc}
\hline
\hline
Emission & $S_{330}$ & $S_{1.5}$ & Size & $\alpha$ \\
~&  (mJy) & (mJy) & (arcsec$^2$) & ($S_{\nu} \propto \nu^{-\alpha}$)\\   
\hline
~&~&~&~\\
Nuclear & $<$9.3 & $13.5 \pm 1$ & -  & $\leq -0.25$\\
Diffuse & $990 \pm 50$ & $29 \pm 2$ & $135 \times 128$ & $2.37 \pm 0.05$ \\
\hline
\label{Gitti:risradio.tab}
\end{tabular}
\end{center}
\end{table}

The extended radio 
source observed at the center of A2626 is characterized by amorphous 
morphology, lack of polarized flux and very steep spectrum which
steepens with distance from the center.  
Finally, the morphology of the diffuse radio emission
is similar to that of the cooling flow region \citep{rizza00}. 
These results indicate that the diffuse radio source may be 
classified as a mini--halo.


\subsection{Model application}

In order to test the prediction of the model, 
we have calculated for both Perseus and A2626
the following expected observable properties:

\textit{total spectrum}:
the total synchrotron spectrum is obtained by integrating 
the synchrotron 
emissivity from an electron population with the energy
distribution given by Eq. \ref{Gitti:ennegamma.eq} over the cluster volume,
assumed to be spherical (we notice that the resulting spectrum in this model 
is weakly dependent on the radius of the sphere, and thus the 
assumption of spherical symmetry is reliable);

\textit{brightness profile}: 
the surface brightness profile expected by the model is obtained
by integrating along the line of sight 
the synchrotron emissivity at a particular frequency;

\textit{radial spectral steepening}:
at varying projected radius, we obtain the surface brightness at 
different frequencies and then compute the spectral index.

The expected brightness profile and radial spectral steepening 
in the model are compared to the observed profiles along the radial 
direction indicated in Fig. \ref{Gitti:perseus.fig} 
for Perseus and in Fig. \ref{Gitti:a2626halo.fig} for A2626.

\begin{figure}
\plotone{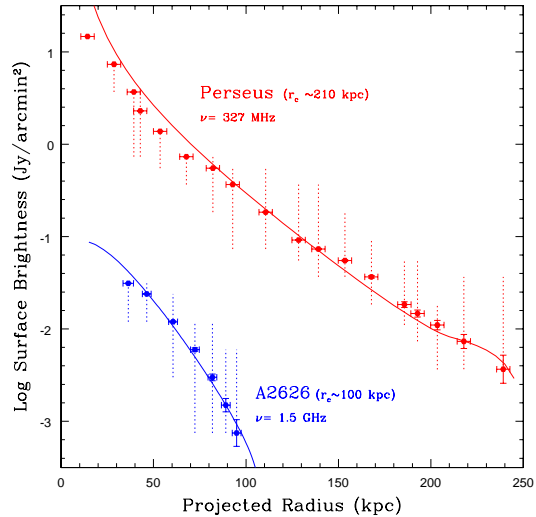}
\figcaption{
Fit to the surface brightness profile at 327 MHz for Perseus
(in red) and 1.5 GHz for A2626 (in blue). 
Vertical bars represent the deviations from the spherical symmetry 
of the diffuse radio emission in other directions in the cluster with respect 
to the one we have considered (see Fig. \ref{Gitti:perseus.fig} for Perseus
and Fig. \ref{Gitti:a2626halo.fig} for A2626). 
\label{Gitti:brillanza.fig}}
\end{figure}

\begin{figure}
\plotone{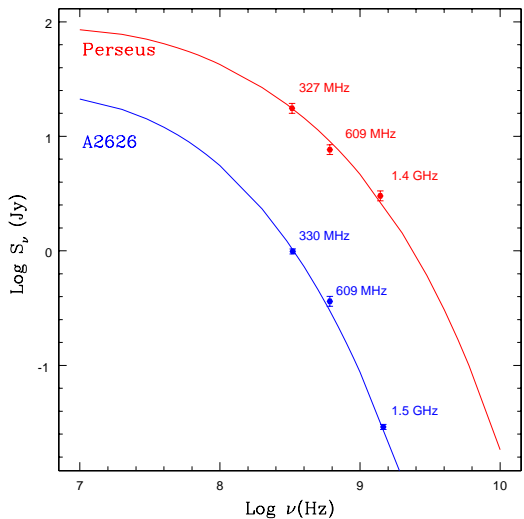}
\figcaption{
Fit to the total spectrum of the synchrotron emission for Perseus
(in red) and A2626 (in blue).
Radio data for Perseus are taken from \citet{sijbring93}.
Radio data for A2626 are taken from Tab. \ref{Gitti:risradio.tab}, 
with the flux density at 609 MHz obtained by subtracting 
the estimated core emission from the total flux given by \citet{roland85}.
\label{Gitti:spettro.fig}}
\end{figure}

\begin{figure}
\plotone{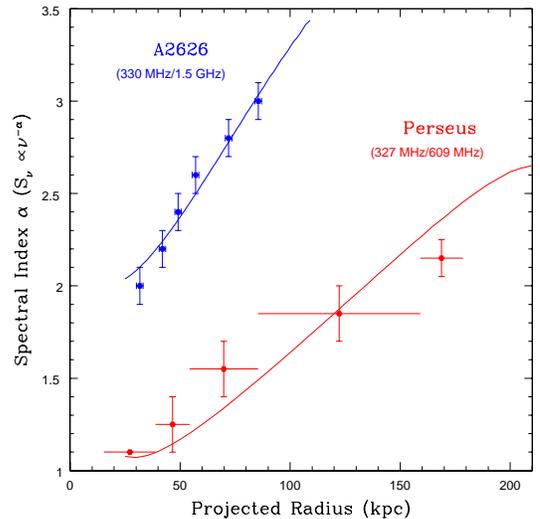}
\figcaption{
Fit to the spectral steepening with distance between 
$\nu=327$ MHz and $\nu=609$ MHz for Perseus (in red),
and $\nu=330$ MHz and $\nu = 1.5$ GHz for A2626 (in blue).
Radio data for Perseus are taken from Fig. \ref{Gitti:perseus_irr.fig}.
Radio data for A2626 are taken from Fig. \ref{Gitti:a2626_irr.fig}.
\label{Gitti:irr.fig}}
\end{figure}

We found good results by assuming an isotropic compression of the 
magnetic field: in particular we notice that,
by reproducing the observed brightness profile,
the model is able to match the integrated spectrum and the radial spectral
steepening as well. 
In Fig. \ref{Gitti:brillanza.fig}, \ref{Gitti:spettro.fig} and 
\ref{Gitti:irr.fig} we show the representative fits 
to the surface brightness profile, total spectrum and 
radial spectral steepening for Perseus (in red) and A2626 (in blue).

\begin{table*}
\begin{center}
\caption{Data and model results for Perseus and A2626}
\begin{tabular}{cc|c|c|c||c|c||c|c|c||c|c|c|c|}
\multicolumn{1}{c}{~} & \multicolumn{4}{c}{X--RAY DATA} & \multicolumn{2}{c}{RADIO DATA} & \multicolumn{3}{c}{MODEL PARAMETERS} & \multicolumn{4}{c}{MODEL RESULTS}\\
\hline
\hline
\multicolumn{1}{|c||}{Cluster} &  $\dot{M}$ & $r_{\rm c}$ & $kT$ & $P_{\rm CF}$ & $P_{1.5}$ & $r_{\rm mh}$  & $B_{\rm c}$ & $l_{\rm c}/[\delta B_{\rm c} / B_{\rm c}]^2$ & s & $\gamma_{\rm b,c}$ & $E_{\rm e}$ & $ N_{\rm e}$ & $\epsilon$\\
\multicolumn{1}{|c||}{~} & ($\mbox{M}_{\odot} \mbox{ yr}^{-1}$) & (kpc) & (keV) & (erg s$^{-1}$) & (W Hz$^{-1}$) &(kpc)&($\mu \mbox{G}$) & (pc) &~ & ~ & (erg) &~ & (\%)\\
\hline
\multicolumn{1}{|c||}{~}&~&~&~&~&~&~&~&~&~&~&~&~&~\\
\multicolumn{1}{|c||}{Perseus} & $519^{+3}_{-17}$ & $210^{+100}_{-20}$ & $6.33^{+0.21}_{-0.18}$ & $2.6 \times 10^{44}$ & $4.4 \times 10^{24}$ & $\sim 300$ & 0.9 - 1.2 &  35 - 60  & $\sim 2 $ & 1600 & $1.6 \times 10^{58} $ & $1.6 \times 10^{61} $ & $ 0.34$\\
\multicolumn{1}{|c||}{A2626} &  $53^{+36}_{-30}$ & $114^{+50}_{-59}$ & $3.1^{+0.5}_{-1}$ & $1.2\times10^{43}$ & $4.3 \times 10^{23}$ & $\sim 100$ & 1.2 - 1.6 & 120 - 180  & $\sim -0.5 $ & 1100 & $2.2 \times 10^{57} $ & $3.2 \times 10^{60} $ & $ 0.69$\\  
\hline
\label{Gitti:confr_ris.tab}
\end{tabular}
\end{center}
\textit{Notes}: 
Columns 2 and 3 list the cooling flow parameters, taken from \citet{ettori98} 
with \textit{ROSAT} PSPC for Perseus and from 
\citet{white97} with \textit{Einstein} IPC for A2626.
Column 4 lists the average temperature of the ICM, taken from \citet{allen92}
with \textit{Ginga} for Perseus and from \citet{white97}
with \textit{Einstein} IPC for A2626.
Column 5 lists the power supplied by the cooling flow as estimated from
X--ray data (see text).
Columns 6 and 7 list the physical properties of the mini--halo: total 
power at 1.5 GHz and radius; 
radio data for Perseus are taken from \citet{sijbring93}.
Columns 8, 9 and 10 list the parameters of the model derived by fitting 
all the observational constraints. Column 11, 12, 13 list the physical 
properties derived by the model: respectively, the break energy of the 
electron spectrum at $r_{\rm c}$, the energetics associated 
with the electrons re--accelerated in the cooling flow region and
their total number.
Column 14 lists the efficiency $\epsilon$ required for 
re--accelerating the relativistic electrons, given in percentage of 
$P_{\rm CF}$ taken from column 5.
See text for details.
\end{table*}

\subsection{Discussion}

In this Section we discuss the physical implications derived by applying 
GBS's model to Perseus and A2626.
For completeness the X--ray and radio data as well as model results
for these clusters are listed in Table \ref{Gitti:confr_ris.tab}.

It is worth pointing out that GBS's model is able to reproduce all the
observational constraints of Perseus and A2626 for physically--meaningful 
values of the parameters (Table \ref{Gitti:confr_ris.tab}). 
First of all we found that the range of values obtained for $B_{\rm c}$
is in agreement with the measurements of magnetic field strengths in the ICM
(Carilli \& Taylor 2002 and references therein). 
We also point out that in the general theory of turbulent plasma one can 
calculate the wavelength which carries most of the turbulent energy 
in a spectrum of Alfv\'en waves \citep[e.g.,][]{tsytovich72}.
When applied to the case of the ICM, with standard values of the physical
parameters, it gives results of the order of tens to hundreds
pc (GBS), thus the $l_{\rm c}/[\delta B_{\rm c} / B_{\rm c}]^2$ values
found in the model are reasonable.

The energetics associated with the population of electrons re--accelerated
in the cooling flow region can be estimated as:
\begin{equation}
\label{energetica.eq}
E_{\rm e} \approx \frac{\pi r_{\rm c}^3 \, {\rm e}^2 \, m_{\rm e} c^2 
N(\gamma_{\rm b})_{\rm c} \,  \gamma_{\rm b,c}^2}{(3-s)} 
\end{equation}
where $N(\gamma_{\rm b})_{\rm c}$ is the number density (per $\gamma$ unit) 
of electrons with energy $\gamma_{\rm b} m_{\rm e} c^2$ at $r_{\rm c}$.
The total number of relativistic electrons can be derived from 
the energetics as:
$N_{\rm e} \sim 4 E_{\rm e}/(3 m_{\rm e} c^2 \, \gamma_{\rm b,c})$.

It is worth noticing that both the energetics and the number of 
re--accelerated electrons estimated for A2626 are 
about one order of magnitude smaller than those found in Perseus.
This is consistent with the fact that the radio power of the mini--halo 
in A2626 is about one order of magnitude smaller than that in Perseus
(see column 6 of Table \ref{Gitti:confr_ris.tab}).

The power $ P_{\rm e}$ 
necessary to re--accelerate the emitting electrons, given by
the minimum energy one must supply to 
balance the radiative losses of these electrons
($P_{\rm e} \approx m_{\rm e} c^2 \beta \, 
\gamma_{\rm b,c}^2 \cdot  N_{\rm e}$,
where $\beta$ is the same as in Eq. \ref{Gitti:gammapunto.eq})
should be significantly smaller than
the power supplied by the cooling flow.
The maximum possible power $P_{\rm CF}$ which can be extracted from the 
cooling flow itself can be estimated assuming a standard cooling
flow model and it corresponds to the $p \, dV/dt$ work done on the gas 
per unit time as it enters $r_{\rm c}$ : 
$P_{\rm CF} = p_{\rm c} \cdot 4 \pi r_{\rm c}^2 \cdot v_{\rm F,c}
\sim  2/5 L_{\rm cool}$ (e.g., \citeauthor{fabian94} 1994, 
$L_{\rm cool}$ being the luminosity associated with the cooling region).

For both the clusters studied here, one finds that only a small fraction 
of $P_{\rm CF}$ should be converted into electron re--acceleration
(see column 14 of Table \ref{Gitti:confr_ris.tab}),
and thus we conclude that processes powered by 
the cooling flow itself are able to provide sufficient energy 
to power the radio mini--halos in Perseus and A2626.  

For a more detailed discussion see \citet{gitti03}.


\section{Observational properties of mini--halos}
\label{Gitti:minihalos.sec}

We collected data from the literature and
studied the observational properties of the
population of radio mini--halos in relation with those of host clusters.
The clusters in the sample were selected based on the presence of both a
cooling flow and a diffuse radio emission with no direct association
with the central radio source. 
In particular, since GBS's model assumes a connection between the origin of 
the synchrotron emission and the cooling flow, to be conservative 
we selected those clusters where the size of the diffuse radio emission
is comparable to the cooling radius. 

\begin{figure}
\plotone{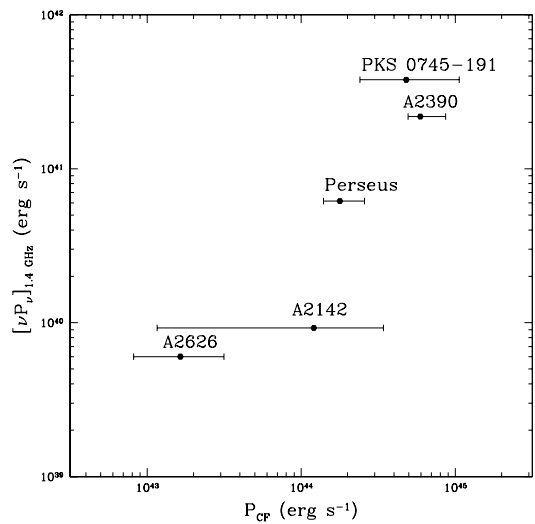}
\figcaption{
Integrated radio power at 1.4 GHz, $\left[ \nu P_{\nu} \right]_{\rm 1.4 GHz}$,
vs. cooling flow power, $P_{\rm CF} = \dot{M} k T/\mu m_{\rm p}$, 
for the mini--halo clusters in Tab. \ref{Gitti:minialoni.tab}.
\label{Gitti:correlazione.fig}}
\end{figure}

Relevant X--ray and radio data are reported in Tab. \ref{Gitti:minialoni.tab},
along with references, while in Fig. \ref{Gitti:correlazione.fig} we 
plotted the radio power at 1.4 GHz of the mini--halos 
versus the maximum power of cooling flows.
A general trend is found, with the strongest radio mini--halos 
associated with the most powerful cooling flows.

\begin{table}
\begin{center}
\caption{Observational data for mini--halos}
\begin{tabular}{lcccc}
\hline
\hline
Cluster & $z$ & $\dot{M}$ & $kT$ & $P_{1.4}$ \\
~&~&  ($\mbox{M}_{\odot} \mbox{ yr}^{-1}$) & (keV) & (W Hz$^{-1}$)\\  
\hline
~&~&~&~\\
PKS 0745-191 & 0.1028 & $579^{+399}_{-215}$ & $8.3^{+0.5}_{-5.8}$ & $2.7 \times 10^{25}$\\
Abell 2390 & 0.2320 & $625^{+75 }_{-75 }$ & $9.5^{+1.3}_{-3.4}$ & $1.5 \times 10^{25}$\\
Perseus & 0.0183 & $283^{+14 }_{-12}$ & $6.3^{+1.5}_{-2.3}$ & $4.4 \times 10^{24}$\\
Abell 2142 & 0.0890 & $106^{+248}_{-106}$ & $11.4^{+0.8}_{-3.2}$ & $6.6 \times 10^{23}$\\
Abell 2626 & 0.0604  & $ 53^{+36 }_{-30}$ & $3.1^{+0.5}_{-1.0}$ & $4.3 \times 10^{23}$\\
\hline
\label{Gitti:minialoni.tab}
\end{tabular}
\end{center}
\textit{References}:
{\bf X--ray data:} 
Perseus, A2142, A2626: \citet{white97} with {\it Einstein} IPC;
PKS 0745-191: \citet{white97} with {\it Einstein} HRI; 
A2390: $\dot{M}$ from \citet{bohringer98} with {\it ROSAT} PSPC, 
$kT$ from \citet{white97} with {\it Einstein} IPC.
{\bf Radio data:} 
PKS 0745-191: \citet{baum91};
A2390: \citet{bacchi03}; Perseus: \citet{sijbring93};
A2142: \citet{giovannini00}; A2626: \citet{gitti03}.
\end{table}

We notice that the maximum powers which can be extracted from
cooling flows are orders of magnitude 
larger than the integrated radio powers (see Fig. 
\ref{Gitti:correlazione.fig}), 
in qualitative agreement with the very low efficiencies 
calculated in the model (see column 14 of Table \ref{Gitti:confr_ris.tab}).

As already discussed in Sect. \ref{Gitti:mymodel.sec},
it is worth noticing that new {\it Chandra} and {\it XMM--Newton} 
results obtained for a limited number of objects hint to an 
overestimate of $\dot{M}$ derived by earlier observations: in particular, 
the consensus reached in these studies
(e.g., \citeauthor{mcnamara00} 2000; \citeauthor{peterson01} 2001;
\citeauthor{david01} 2001; \citeauthor{molendi01} 2001; 
\citeauthor{bohringer01} 2001; \citeauthor{peterson03} 2003) is
that the spectroscopically--derived cooling rates are a factor $\sim$ 5-10
less than earlier {\it ROSAT} and {\it Einstein} values
\citep[e.g.,][]{fabian03}. 
This factor seems to be similar for all clusters in a large range of 
$\dot{M}$, giving a systematic effect that will not spoil the correlation 
reported in Fig. \ref{Gitti:correlazione.fig}. 

We can conclude that even by considering reduced spectroscopically--derived  
$\dot{M}$, the cooling flow powers are well above the radio powers emitted 
by mini--halos, therefore
even if the cooling flow was partially re--heated and
stopped, the basic idea and results of the model would not change.


\acknowledgements

M.G. would like to thank F. Brighenti for stimulating discussion during the 
Conference and S. Dall'Osso for his constant help and insightful comments.
M.G. and G.B. acknowledge partial support from CNR grant CNRG00CF0A.
This work was partly supported by the Italian Ministry for University
and Research (MIUR) under grant Cofin 2001-02-8773 and by
the Austrian Science Foundation FWF under grant P15868.


\end{document}